# The Capacity Region of the Degraded Finite-State Broadcast Channel

Ron Dabora and Andrea Goldsmith
Dept. of Electrical Engineering, Stanford University, Stanford, CA

*Abstract*— We consider the discrete, time-varying broadcast channel with memory under the assumption that the channel states belong to a set of finite cardinality. We first define the *physically degraded* finite-state broadcast channel for which we derive the capacity region. We then define the *stochastically degraded* finite-state broadcast channel and derive the capacity region for this scenario as well. In both scenarios we consider the non-indecomposable finite-state channel as well as the indecomposable one.

## I. INTRODUCTION

The broadcast channel (BC) was introduced by Cover in 1972. In this scenario a single sender transmits three messages, one common and two private, to two receivers over a channel defined by $\{\mathcal{X}, p(y,z|x), \mathcal{Y} \times \mathcal{Z}\}$. Here, $X$ is the channel input from the transmitter, $Y$ is the channel output at Rx$_1$ and $Z$ is the channel output at Rx$_2$. In the years following its introduction the study of the BC focused on memoryless scenarios, i.e., when the probability of a block of $n$ transmissions is given by $p(y^n, z^n|x^n) = \prod_{i=1}^{n} p(y_i, z_i|x_i)$. In recent years, models of time-varying broadcast channels with memory have attracted a lot of attention, especially Gaussian BCs. This was motivated by the proliferation of mobile communications, for which the channel is subject to time-varying correlated fading. The correlation of the fading process introduces memory in the BC. The fading BC is one instance of the general BC with channel states. While fading BCs have received considerable attention, discrete, time-varying BCs with channel states have not been well studied. A notable exception is the degraded arbitrarily varying BC (DAVBC) considered in [2] and [3]. In [2] DAVBCs with causal and non-causal side information at the transmitter were considered. The states are assumed i.i.d. and the channel is memoryless: $p(y^n, z^n|x^n, s^n) = \prod_{i=1}^{n} p(y_i, z_i|x_i, s_i)$. In [3], the capacity region for DAVBCs with causal side information at the transmitter and non-causal side information at the good receiver was derived. In [3] the state distribution is general and is not subject to the i.i.d. restriction, but the channel outputs, given the states and the channel inputs are again memoryless. The general, discrete BC with i.i.d. states non-causally known at the transmitter was considered in [4].

The arbitrarily varying channel (AVC) is one model for a time-varying channel with states. It models a memoryless channel whose law varies in time in an arbitrary manner. The state transitions are independent of the channel inputs and outputs. In this work we study the discrete time-varying BC with memory in the framework of finite-state channels (FSCs).

In contrast to the AVC, in the FSC both the channel output and the current state depend on both the channel input and the previous state.

The finite-state channel model was used to model point-to-point channel variations as early as 1953 [1]. This channel is characterized by the distribution $p(y,s|x,s')$ where $S$ is the current state and $S'$ is the previous state. For a block of $n$ transmissions, the p.m.f. at the $i$'th symbol time satisfies

$$p(y_i, s_i|x^i, s^{i-1}, y^{i-1}, s_0) = p(y_i, s_i|x_i, s_{i-1}), \quad (1)$$

where $s_0$ is the state of the channel when transmission began. Equation (1) implies that $S_{i-1}$ contains all the history information for time $i$. Recently, the finite-state multiple-access channel was studied in [6]. This scenario is characterized by the channel distribution $p(y,s|x_1, x_2, s')$, and the work in [6] also considered the effect of feedback on the rates.

In the present work we study the finite-state broadcast channel (FSBC). Here, the channel from the transmitter to the receivers is governed by a state sequence that depends on the channel inputs, outputs and previous states. The way these symbols interact with each other is captured by the transition function $p(y,z,s|x,s')$.

*Main Contributions and Organization*

In this paper we consider for the first time the capacity of the FSBC. Here, there is a unique aspect not encountered in the point-to-point and the MAC counterparts, namely the application of superposition coding to the FSC. We initially define the physically degraded FSBC and find the capacity region of this scenario. We then define the stochastically degraded FSBC and give examples of communication scenarios represented by this model. We derive the capacity region for this channel as well.

The rest of this paper is organized as follows: Section II introduces the channel model. Section III presents a summary of the results together with a discussion. Lastly, Section IV outlines the proof of the capacity region for the physically degraded FSBC.

## II. CHANNEL MODEL AND DEFINITIONS

First, a word about notation. In the following we denote random variables with upper case letters, e.g. $X, Y$, and their realizations with lower case letters $x, y$. A random variable (RV) $X$ takes values in a set $\mathcal{X}$. We use $||\mathcal{X}||$ to denote the cardinality of a finite, discrete set $\mathcal{X}$, $\mathcal{X}^n$ to denote the $n$-fold Cartesian product of $\mathcal{X}$, and $p_X(x)$ to denote the probability mass function (p.m.f.) of a discrete RV $X$ on $\mathcal{X}$. For brevity we may omit the subscript $X$ when it is obvious from the context. We use $p_{X|Y}(x|y)$ to denote the conditional p.m.f. of $X$ given $Y$. We denote vectors with boldface letters, e.g. $\mathbf{x}, \mathbf{y}$;

The authors are with the Wireless Systems Lab, Department of Electrical Engineering, Stanford University, Stanford, CA 94305. Email: {ron,andrea}@wsl.stanford.edu. This work was supported in part by the DARPA ITMANET program under grant 1105741-1-TFIND.

the $i$'th element of a vector $\mathbf{x}$ is denoted with $x_i$ and we use $x_i^j$ where $i < j$ to denote the vector $(x_i, x_{i+1}, ..., x_{j-1}, x_j)$; $x^j$ is short form notation for $x_1^j$, and $\mathbf{x} \equiv x^n$. A vector of $n$ random variables is denoted by $X^n$, and similarly we define $X_i^j \triangleq (X_i, X_{i+1}, ..., X_{j-1}, X_j)$ for $i < j$. We use $H(\cdot)$ to denote the entropy of a discrete random variable and $I(\cdot; \cdot)$ to denote the mutual information between two random variables, as defined in [7, Chapter 2]. $I(\cdot; \cdot)_q$ denotes the mutual information evaluated with a p.m.f. $q$ on the random variables. Finally, $co\,\mathcal{R}$ denotes the convex hull of the set $\mathcal{R}$.

*Definition 1:* The *discrete, finite-state broadcast channel* is defined by the triplet $\{\mathcal{X} \times \mathcal{S}, p(y, z, s|x, s'), \mathcal{Y} \times \mathcal{Z} \times \mathcal{S}\}$ where $X$ is the input symbol, $Y$ and $Z$ are the output symbols, $S'$ is the state of the channel at the end of the previous symbol transmission and $S$ is the state of the channel at the end of the current symbol transmission. $\mathcal{S}, \mathcal{X}, \mathcal{Y}$ and $\mathcal{Z}$ are discrete alphabets of finite cardinalities. The p.m.f of a block of $n$ transmissions is

$$p(y^n, z^n, s^n, x^n | s_0)$$
$$= \prod_{i=1}^{n} p(y_i, z_i, s_i, x_i | y^{i-1}, z^{i-1}, s^{i-1}, x^{i-1}, s_0)$$
$$= \prod_{i=1}^{n} p(x_i | x^{i-1}) p(y_i, z_i, s_i | y^{i-1}, z^{i-1}, s^{i-1}, x^i, s_0)$$
$$\stackrel{(a)}{=} p(x^n) \prod_{i=1}^{n} p(y_i, z_i, s_i | x_i, s_{i-1}), \qquad (2)$$

where $s_0$ is the initial channel state. Here (a) captures the fact that given $S_{i-1}$, the symbols at time $i$ are independent of the past.

*Definition 2:* The FSBC is called *physically degraded* if its p.m.f. satisfies

$$p(y_i|x^i, y^{i-1}, z^{i-1}, s_0) = p(y_i|x^i, y^{i-1}, s_0), \qquad (3a)$$
$$p(z_i|x^i, y^i, z^{i-1}, s_0) = p(z_i|y^i, z^{i-1}, s_0). \qquad (3b)$$

Condition (3a) captures the intuitive notion of degradedness, namely that $Z^{i-1}$ is a degraded version of $Y^{i-1}$, thus it does not add information when $Y^{i-1}$ is given. Note that in the memoryless case this condition is not necessary as, given $X_i$, $Y_i$ is independent of the history. Condition (3b) follows from the standard notion of degradedness.

Using conditions (3a) and (3b) we obtain (when $p(y^n, x^n|s_0) > 0$)

$$p(z^n|y^n, x^n, s_0)$$
$$= \frac{p(z^n, y^n, x^n|s_0)}{p(y^n, x^n|s_0)}$$
$$= \frac{\prod_{i=1}^{n} p(z_i, y_i, x_i | z^{i-1}, y^{i-1}, x^{i-1}, s_0)}{\prod_{i=1}^{n} p(y_i, x_i | y^{i-1}, x^{i-1}, s_0)}$$
$$= \frac{\prod_{i=1}^{n} p(x_i | z^{i-1}, y^{i-1}, x^{i-1}) \prod_{i=1}^{n} p(z_i, y_i | z^{i-1}, y^{i-1}, x^i, s_0)}{\prod_{i=1}^{n} p(x_i | y^{i-1}, x^{i-1}) \prod_{i=1}^{n} p(y_i | y^{i-1}, x^i, s_0)}$$
$$\stackrel{(a)}{=} \frac{\prod_{i=1}^{n} p(x_i | x^{i-1}) \prod_{i=1}^{n} p(z_i, y_i | z^{i-1}, y^{i-1}, x^i, s_0)}{\prod_{i=1}^{n} p(x_i | x^{i-1}) \prod_{i=1}^{n} p(y_i | y^{i-1}, x^i, s_0)}$$
$$\stackrel{(b)}{=} \frac{\prod_{i=1}^{n} p(y_i | y^{i-1}, x^i, s_0) \prod_{i=1}^{n} p(z_i | z^{i-1}, y^i, x^i, s_0)}{\prod_{i=1}^{n} p(y_i | y^{i-1}, x^i, s_0)}$$
$$\stackrel{(c)}{=} \prod_{i=1}^{n} p(z_i | z^{i-1}, y^i, s_0), \qquad (4)$$

where (a) is because there is no feedback, (b) follows from (3a) and (c) follows from (3b). We conclude that when (3) holds, $p(z^n|y^n, x^n, s_0) = p(z^n|y^n, s_0)$. Hence,

$$p(y^n, z^n|x^n, s_0) = p(y^n|x^n, s_0)p(z^n|y^n, s_0). \qquad (5)$$

Note that (4) shows how to obtain $p(z^n|y^n, x^n, s_0)$ in a causal manner. Also note that $Z^n$ is a degraded version of $Y^n$ but still depends on the state sequence (i.e. degradedness does not eliminate the memory). A special case of the physically degraded FSBC occurs when in (3b) it holds that $p(z_i|x^i, y^i, z^{i-1}, s_0) = p(z_i|y_i)$. Hence,

$$p(z^n|y^n, x^n, s_0) = p(z^n|y^n) = \prod_{i=1}^{n} p(z_i|y_i). \qquad (6)$$

Equation (6) is similar to the definition of degradedness for the DAVBC used in [2].

*Definition 3:* The FSBC is called *stochastically degraded* if there exists a p.m.f. $\tilde{p}(z|y)$ such that

$$p(z, s|x, s') = \sum_{\mathcal{Y}} p(y, s|x, s') p(z|y, s, x, s')$$
$$= \sum_{\mathcal{Y}} p(y, s|x, s') \tilde{p}(z|y). \qquad (7)$$

Note that when (7) holds then

$$p(z^n|x^n, s_0) = \sum_{\mathcal{S}^n} p(z^n, s^n | x^n, s_0)$$
$$\stackrel{(a)}{=} \sum_{\mathcal{S}^n} \prod_{i=1}^{n} p(z_i, s_i | x_i, s_{i-1})$$
$$= \sum_{\mathcal{S}^n} \prod_{i=1}^{n} \sum_{y_i \in \mathcal{Y}} p(y_i, s_i | x_i, s_{i-1}) \tilde{p}(z_i|y_i)$$
$$= \sum_{\mathcal{S}^n} \sum_{\mathcal{Y}^n} \prod_{i=1}^{n} p(y_i, s_i | x_i, s_{i-1}) \tilde{p}(z_i|y_i)$$
$$\stackrel{(b)}{=} \sum_{\mathcal{S}^n} \sum_{\mathcal{Y}^n} p(y^n, s^n | x^n, s_0) \prod_{i=1}^{n} \tilde{p}(z_i|y_i)$$
$$= \sum_{\mathcal{Y}^n} p(y^n | x^n, s_0) \prod_{i=1}^{n} \tilde{p}(z_i|y_i), \qquad (8)$$

where (a) and (b) follow from (2).

Definition 3 does not constitute only a mathematical convenience, but represents a physical scenario. For example, consider a scenario in which a base station transmits to two mobile units, located approximately on the same line-of-sight from the base station (BS), as indicated by the dashed line in Figure 1. Let the BS transmit a BPSK signal and let the received signals be subject to additive Gaussian thermal noise due to the receivers' front-ends. When decoding at the receivers takes place after a hard threshold at zero, the resulting scenario is the binary symmetric broadcast channel (BSBC). Denote the situation where there is no traffic on the road between the BS and the mobiles as state $A$. Let the channel BS–Rx$_1$ have a crossover probability $\epsilon_1(A) = 0.1$ and the channel BS–Rx$_2$ have a crossover probability $\epsilon_2(A) = 0.15$. This can be represented as a stochastically degraded BC with a degrading channel whose crossover probability is

$$\epsilon_{12}(A) = \frac{\epsilon_2 - \epsilon_1}{1 - 2\epsilon_1} = 0.0625.$$

Assume that on occasions, a car passes on the road between the BS and the mobiles. This causes attenuation in both channels simultaneously. Call this state $B$ and let $\epsilon_1(B) = 0.18$ and $\epsilon_2(B) = 0.22$. Again we have $\epsilon_{12}(B) = 0.0625$[1]. Hence, the degrading channel is the same for both states, irrespective of the state sequence (in this example the state sequence represents the traffic pattern, and is not an independent sequence). This satisfies condition (8).

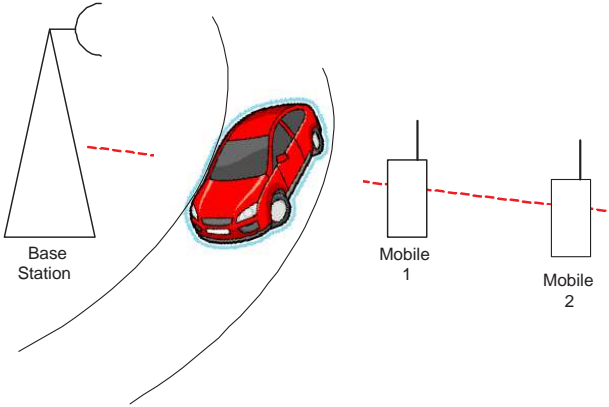

Fig. 1. A degraded FSBC scenario: the mobile units are located on the same line-of-sight from the base-station (indicated by the dashed line). Passing cars affect the channels to both mobile units simultaneously.

More generally, we can define a set of states for this scenario, e.g. $\mathcal{S} = \{1, 2, ..., K\}$, with $\mathcal{Y} = \mathcal{Z} = \{0, 1\}$ and

$$p(z_i, s_i | y_i, s_{i-1}) = p(s_i | s_{i-1}) p(z_i | y_i, s_i)$$
$$p(z|y, s = k) = \begin{cases} \epsilon_{12}(k) & , z = 1 - y \\ 1 - \epsilon_{12}(k) & , z = y \end{cases},$$

$\epsilon_{12}(k) \in (0, 0.5), k \in \mathcal{S}$. This results in a collection of physically degraded BSBCs that can give more flexibility in modeling the scenario of Figure 1, as the degrading channel may depend on the state. However, for this reason, this model does not satisfy our definition of stochastic degradedness in Definition 3.

*Definition 4:* (see [5, Section 4.6]) The FSBC is called *indecomposable* if for every $\epsilon > 0$ there exists $N_0(\epsilon)$ such that for all $n > N_0(\epsilon)$, $|p(s_n | \mathbf{x}, s_0) - p(s_n | \mathbf{x}, s'_0)| < \epsilon$, for all $s_n$, $\mathbf{x}$, and initial states $s_0$ and $s'_0$.

*Definition 5:* An $(R_0, R_1, R_2, n)$ deterministic code for the FSBC consists of three message sets, $\mathcal{M}_0 = \{1, 2, ..., 2^{nR_0}\}$, $\mathcal{M}_1 = \{1, 2, ..., 2^{nR_1}\}$ and $\mathcal{M}_2 = \{1, 2, ..., 2^{nR_2}\}$, and three mappings $(f, g_\mathbf{y}, g_\mathbf{z})$ such that

$$f : \mathcal{M}_0 \times \mathcal{M}_1 \times \mathcal{M}_2 \mapsto \mathcal{X}^n \qquad (9)$$

is the encoder and

$$g_\mathbf{y} : \mathcal{Y}^n \mapsto \mathcal{M}_0 \times \mathcal{M}_1,$$
$$g_\mathbf{z} : \mathcal{Z}^n \mapsto \mathcal{M}_0 \times \mathcal{M}_2,$$

are the decoders. Here, $\mathcal{M}_0$ is the set of common messages and $\mathcal{M}_1$ and $\mathcal{M}_2$ are the sets of private messages to $\text{Rx}_1$ and $\text{Rx}_2$ respectively.

---
[1] The scenario parameters assumed in this example are: Two-ray propagation model, Rx decoding scheme is maximum-likelihood, Base station Tx power = 30 dBm, Base station antenna gain = 10 dBi, Rx antenna gain = 0 dBi, Rx noise floor = $-90$ dBm, Base station antenna height = 10 m, Rx antenna height = 1.5 m, BS–Rx$_1$ distance = 7.2 Km and BS–Rx$_2$ distance = 8 Km. We also assume a passing car increases the path attenuation by 3 dB.

Note that we assume *no knowledge of the states at the transmitter and receivers*.

*Definition 6:* The *average probability of error* of a code for the FSBC is given by $P_e^{(n)} = \max_{s_0 \in \mathcal{S}} P_e^{(n)}(s_0)$, where,

$$P_e^{(n)}(s_0) = \Pr\big(g_\mathbf{y}(Y^n) \neq (M_0, M_1) \text{ or}$$
$$g_\mathbf{z}(Z^n) \neq (M_0, M_2) | s_0\big),$$

where each of the messages $M_0 \in \mathcal{M}_0$, $M_1 \in \mathcal{M}_1$ and $M_2 \in \mathcal{M}_2$ is selected independently and uniformly.

*Definition 7:* A rate triplet $(R_0, R_1, R_2)$ is called *achievable* for the FSBC if for every $\epsilon > 0$ and $\delta > 0$ there exists an $n(\epsilon, \delta)$ such that for all $n > n(\epsilon, \delta)$ an $(R_0 - \delta, R_1 - \delta, R_2 - \delta, n)$ code with $P_e^{(n)} \leq \epsilon$ can be constructed.

*Definition 8:* The *capacity region* of the FSBC is the convex hull of all achievable rate triplets.

### III. MAIN RESULTS AND DISCUSSION

Define first

$$R_{1,n}(p, s_0) \triangleq \frac{1}{n} I(X^n; Y^n | U^n, s_0)_p - \frac{\log_2 ||\mathcal{S}||}{n}$$
$$R_{2,n}(p, s_0) \triangleq \frac{1}{n} I(U^n; Z^n | s_0)_p - \frac{\log_2 ||\mathcal{S}||}{n}.$$

The main result is stated in the following theorem, whose proof is outlined in Section IV:

*Theorem 1:* Let $\mathcal{Q}_n$ be the set of all joint distributions on $(\times_{i=1}^n \mathcal{U}_i, \mathcal{X}^n)$ such that the cardinality of the random vector $U^n$ is bounded by $||\times_{i=1}^n \mathcal{U}_i|| \leq \min\{||\mathcal{X}||, ||\mathcal{Y}||, ||\mathcal{Z}||\}^n$. For the physically degraded FSBC of Definition 2, define the region $\mathcal{R}_n(s_0)$ as

$$\mathcal{R}_n(s_0) = co \bigcup_{q_n \in \mathcal{Q}_n} \Big\{ (R_0, R_1, R_2) : R_0 \geq 0, R_1 \geq 0, R_2 \geq 0,$$

$$R_1 \leq R_{1,n}(q_n, s_0), \ R_0 + R_2 \leq R_{2,n}(q_n, s_0) \Big\}. (10)$$

*The capacity region of the physically degraded FSBC is given by*

$$\mathcal{C}_{pd} = \lim_{n \to \infty} \bigcap_{s_0 \in \mathcal{S}} \mathcal{R}_n(s_0), \qquad (11)$$

*and the limit exists.*

Since the capacity of the broadcast channel depends only on the conditional marginals $p(y^n | x^n, s_0)$ and $p(z^n | x^n, s_0)$ (see [7, Chapter 14.6]) then the capacity region of the stochastically degraded FSBC is the same as the corresponding physically degraded FSBC:

*Corollary 1:* For the stochastically degraded FSBC of Definition 3, the capacity region is given by Theorem 1 where $p(z|s, y, x, s')$ is replaced by $\tilde{p}(z|y)$ that satisfies equation (7).

When the FSBC is indecomposable, then the effect of the initial state fades away as $n$ increases. Therefore we have the following corollary:

*Corollary 2:* For the indecomposable physically degraded FSBC, the capacity region is given by Theorem 1. For the indecomposable stochastically degraded FSBC, the capacity region is obtained from Corollary 1. In both cases the parameter $s_0$ in $R_{1,n}(q_n, s_0)$ and $R_{2,n}(q_n, s_0)$ and the intersection over $\mathcal{S}$ in the expression for $\mathcal{C}_{pd}$ are omitted.

*Proof outline:* Loosely speaking, the corollary is true since for $n$ large enough the effect of the initial state fades away. Therefore, for asymptotically large $n$ the maximum over all initial states $s_0 \in \mathcal{S}$ equals the minimum.

*Discussion*

First, note that if $\lim_{n \to \infty} \mathcal{R}_n(s_0)$ exists for all $s_0 \in \mathcal{S}$ then the capacity region (11) can be written as

$$\mathcal{C}_{pd} = \lim_{n \to \infty} \bigcap_{s_0 \in \mathcal{S}} \mathcal{R}_n(s_0) \stackrel{(a)}{=} \bigcap_{s_0 \in \mathcal{S}} \lim_{n \to \infty} \mathcal{R}_n(s_0).$$

Here, (a) is permitted because $\mathcal{S}$ is finite. Thus, the capacity region can be viewed as the intersection of all the capacity regions obtained when the initial state is known at the receivers (but not at the transmitter). We also note the following conclusions:

1) Since the limit of the region exists, then as $n$ increases, optimizing the code will result in better performance (*which is not guaranteed when the limits cannot be shown to exist*, consider for example a non-stationary channel with noise that oscillates with time).

2) The codebook structure that achieves capacity is a superposition codebook. This introduces a structural constraint when optimizing the codebook for achieving the maximum rate triplets.

3) The auxiliary RV $U^n$ introduces difficulties mainly in places where we need to rely on the its cardinality. This is because we cannot translate the bound on the cardinality of $U^n$ into a bound on the cardinality of a subset of $U^n$. In particular, we cannot use the cardinality of $U^n$ when deriving the capacity region for the indecomposable FSBC. Moreover, letting $n = m_1 + m_2$, then from Equation (1) we have that

$$p(z^{m_1}, y^{m_1}, s^{m_1} | x^n, s_0) = p(z^{m_1}, y^{m_1}, s^{m_1} | x^{m_1}, s_0).$$

But because $p(x^{m_1} | u^n) \neq p(x^{m_1} | u^{m_1})$ then

$$p(z^{m_1}, y^{m_1}, s^{m_1} | u^n, s_0) \neq p(z^{m_1}, y^{m_1}, s^{m_1} | u^{m_1}, s_0).$$

This is a major difference from the point-to-point and the MAC channels. Consider, for example, the expression

$$\max_{p(u^n, x^n)} \left\{ \max_{s_0 \in \mathcal{S}} \frac{1}{n} I(U^n; Z^n | s_0) + \lambda \max_{s_0' \in \mathcal{S}} \frac{1}{n} I(X^n; Y^n | U^n, s_0') \right\}. \quad (12)$$

While in the MAC and the point-to-point channels the corresponding expressions converge for all channels, for the FSBC (12) can be shown to converge only for the indecomposable case. Therefore, using superposition coding, the channel between $U^n$ and $(Y^n, Z^n)$ is fundamentally different from the channel between $X^n$ and $(Y^n, Z^n)$. This is in contrast also to the discrete, memoryless BC.

## IV. PROOF OUTLINE

In the derivation we focus on the physically degraded FSBC. The derivation requires only that condition (5) holds. In the derivation we shall consider only the two private messages case as the common message can be incorporated by splitting the rate to $Rx_2$ into private and common rates, as in [7, Theorem 14.6.4].

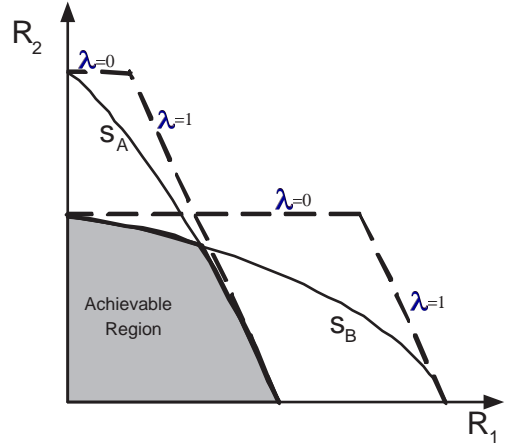

Fig. 2. Lines bounding the achievable regions for the FSBC for initial states $s_A$ and $s_B$, and the resulting region of positive error exponents.

### A. Achievability Theorem

Due to space limitations we omit the details of the achievability proof and give only the conclusion. For complete details see [8]. Define first

$$F_n(\lambda) = \max_{p(u^n, x^n)} \left\{ \min_{s_0 \in \mathcal{S}} R_{2,n}(p, s_0) + \lambda \min_{s_0' \in \mathcal{S}} R_{1,n}(p, s_0') \right\}.$$

Following [9, Section 2], the boundary of the region of positive error exponents for a given $n$ can be written as

$$R_2^n(R_1^n) = \inf_{0 \leq \lambda \leq 1} \left\{ F_n(\lambda) - \lambda R_1 \right\}. \quad (13)$$

This characterization is illustrated in Figure 2.

In the achievability proof we show that for a given $p(u^n, x^n)$, when transmitting at the *positive* rate pair $\left( \min_{s_0' \in \mathcal{S}} R_{1,n}(p, s_0'), \min_{s_0 \in \mathcal{S}} R_{2,n}(p, s_0) \right)$, then the error exponent is positive and bounded away from zero. Hence, the probability of error can be made less than any arbitrary $\epsilon > 0$ by taking a block length $Kn$ with a large enough integer $K$.

Furthermore, in section IV-D we show that the largest region is obtained by taking the limit

$$R_2(R_1) = \inf_{0 \leq \lambda \leq 1} \left\{ \lim_{n \to \infty} F_n(\lambda) - \lambda R_1 \right\}, \quad (14)$$

and that this limit exists and is finite. The fact that the limit exists and is finite implies that we can approach the rates of Theorem 1 arbitrarily close by taking $n$ large enough, thus by Definition 7 these rates are achievable.

Before considering the converse we discuss the cardinality of the auxiliary RV $U^n$, as the evaluation of $R_2^n(R_1^n)$ of (13) depends on the existence of such a bound.

### B. Cardinality Bounds

From the derivation in [9], it follows that maximizing the region $\mathcal{R}_n(s_0)$ of Equation (10) over all joint distributions $p(u^n, x^n)$, can be carried out while the cardinality of the auxiliary random variable $U^n$ is bounded by

$$\| \times_{i=1}^n \mathcal{U}_i \| \leq \min \left\{ \|\mathcal{X}\|, \|\mathcal{Y}\|, \|\mathcal{Z}\| \right\}^n. \quad (15)$$

Now note that from (11), the achievable region for a fixed $n$ is given by the intersection $\bigcap_{s_0 \in \mathcal{S}} \mathcal{R}_n(s_0)$. As for each $\mathcal{R}_n(s_0)$, $s_0 \in \mathcal{S}$ we have the same cardinality bound, then this bound also holds for maximizing the intersection of the regions $\mathcal{R}_n(s_0), s_0 \in \mathcal{S}$.

## C. Converse

*Lemma 1:* If for some $\epsilon > 0$, $\lambda \geq 0$,
$$R_2 + \lambda R_1 > \lim_{n \to \infty} F_n(\lambda) + \epsilon,$$
then there exists a pair of initial states $s_0$ and $s_0'$ such that
$$P_{e2}^{(n)}(s_0)R_2 + \lambda\left(P_{e1}^{(n)}(s_0')R_1\right) > \epsilon - \frac{1}{n}(1+\lambda)(1+\log_2||\mathcal{S}||).$$
The implication of this inequality, as explained in [9, Section 3], is that for $n$ large enough the probability of error $P_e^{(n)}$ cannot be made arbitrarily small outside the region (14).

*Proof:* From Fano's inequality we have that
$$H(M_2|Z^n, s_0) \leq P_{e2}^{(n)}(s_0)nR_2 + 1 \quad (16a)$$
$$H(M_1|Y^n, s_0) \leq P_{e1}^{(n)}(s_0)nR_1 + 1. \quad (16b)$$
Next write
$$\min_{s_0 \in \mathcal{S}} I(M_2; Z^n|s_0) = nR_2 - \max_{s_0 \in \mathcal{S}} H(M_2|Z^n, s_0) \quad (17)$$
$$\min_{s_0' \in \mathcal{S}} I(M_1; Y^n|M_2, s_0') = nR_1 - \max_{s_0' \in \mathcal{S}} H(M_1|Y^n, M_2, s_0')$$
$$\geq nR_1 - \max_{s_0' \in \mathcal{S}} H(M_1|Y^n, s_0'). \quad (18)$$

Now note that
$$I(M_2; Z^n|s_0) = H(Z^n|s_0) - H(Z^n|M_2, s_0)$$
$$= I(U^n; Z^n|s_0), \quad (19)$$
where $U_i = M_2$, $i = 1, 2, ..., n$. We also have
$$I(M_1; Y^n|M_2, s_0') = H(Y^n|M_2, s_0') - H(Y^n|M_1, M_2, s_0')$$
$$\leq H(Y^n|U^n, s_0') - H(Y^n|X^n, U^n, s_0')$$
$$= I(X^n; Y^n|U^n, s_0'), \quad (20)$$
where the definition of $U^n$ satisfies the Markov relationship $U^n|s_0' - X^n|s_0' - Y^n|s_0'$. Combining (19) and (20) we have that for this choice of $U^n$:
$$\min_{s_0 \in \mathcal{S}} I(M_2; Z^n|s_0) + \lambda \min_{s_0' \in \mathcal{S}} I(M_1; Y^n|M_2, s_0')$$
$$\leq \min_{s_0 \in \mathcal{S}} I(U^n; Z^n|s_0) + \lambda \min_{s_0' \in \mathcal{S}} I(X^n; Y^n|U^n, s_0')$$
$$\leq nF_n(\lambda) + (1+\lambda)\log_2||\mathcal{S}||, \quad (21)$$
since $F_n(\lambda)$ is obtained by maximizing over all joint distributions $p(u^n, x^n)$ subject to the cardinality constraint (15), which is also satisfied by our choice of $U^n$. Let $s_{0,n}$ and $s_{0,n}'$ be the maximizing states for $H(M_2|Z^n, s_0)$ and $H(M_1|Y^n, s_0')$ respectively.

Plugging (17) and (18) into (21) yields
$$nR_2 - H(M_2|Z^n, s_{0,n}) + \lambda(nR_1 - H(M_1|Y^n, s_{0,n}'))$$
$$- (1+\lambda)\log_2||\mathcal{S}|| \leq nF_n(\lambda).$$
Thus, $H(M_2|Z^n, s_{0,n}) + \lambda H(M_1|Y^n, s_{0,n}') + (1+\lambda)\log_2||\mathcal{S}|| \geq n(R_2 + \lambda R_1 - F_n(\lambda)) \geq n(R_2 + \lambda R_1 - \lim_{n \to \infty} F_n(\lambda)) > n\epsilon$. Combined with (16), this completes the proof of the lemma. ∎

## D. Convergence

In this subsection we show that $\lim_{n \to \infty} F_n(\lambda)$ exists and is finite for the channel under consideration, when $\lambda \in [0, 1]$.

The proof of convergence extends the arguments in [5, Appendix 4A] to the FSBC. The main difficulty here is the introduction of the auxiliary RV $U^n$ and its interaction with the other RVs, $S^n, X^n, Y^n$ and $Z^n$. We actually show that
$$\lim_{n \to \infty} F_n(\lambda) = \sup_n F_n(\lambda)$$
which implies that the limit exists. Due to its length, the full proof is omitted and only the main points are highlighted.

Let $s_0 = s_0^z(l)$ minimize $\frac{1}{l}I(U^l; Z^l|s_0)$ and $s_0' = s_0^y(l)$ minimize $\frac{1}{l}I(X^l; Y^l|U^l, s_0')$, for the triplet $(q_1(u^l, x^l), s_0^z(l), s_0^y(l))$ that achieves the max-min solution for $F_l(\lambda)$, and let $(q_2(u^m, x^m), s_0^z(m), s_0^y(m))$ achieve the max-min solution $F_m(\lambda)$. Finally, let $s_0^z(n)$ and $s_0^y(n)$ be the states that achieve the max-min solution for $F_n(\lambda)$. We show that $F_n(\lambda)$ is sup-additive, i.e., for every integer $m, l \in [0, n]$ with $n = m + l$ we have
$$nF_n(\lambda) \geq lF_l(\lambda) + mF_m(\lambda).$$

Sup-additivity is verified by breaking the length $n$ expressions into expressions of length $l$ and expressions of length $m$. The critical part here is to consider the length $m$ sequence from $l+1$ to $n$. Here we use the fact that given the initial state the channel is stationary, so $p(Z_{l+1}^n, Y_{l+1}^n|x_{l+1}^n, s_l = s_0) = p(Z_1^m, Y_1^m|x_1^m = x_{l+1}^n, s_0)$. This, combined with the fact the cardinality bound depends only on the length of the sequence, leads to the conclusion that the joint distribution $q_2(u_1^m, x_1^m)$ that maximizes $F_m(\lambda)$ will maximize the segment from $l+1$ to $n$ (i.e. is the maximizing distribution for $(U_{l+1}^n, X_{l+1}^n)$, with the same initial state).

Additionally, both $\frac{1}{n}I(U^n; Z^n|s_0)$ and $\frac{1}{n}I(X^n; Y^n|U^n, s_0')$ are bounded from above, independent of $n$:
$$\frac{1}{n}I(U^n; Z^n|s_0) \leq \log_2||\mathcal{Z}||,$$
since all the $Z_i$'s are defined over the same alphabet $\mathcal{Z}_i \equiv \mathcal{Z}$, and similarly $\frac{1}{n}I(X^n; Y^n|U^n, s_0') \leq \log_2||\mathcal{X}||$. Thus, $F_n(\lambda) \leq \log_2||\mathcal{Z}|| + \lambda\log_2||\mathcal{X}|| < \infty$ for any $\lambda \in [0, 1]$. The fact that $F_n(\lambda)$ is bounded from above independent of $n$ and is also sup-additive implies that $\lim_{n \to \infty} F_n(\lambda)$ exists and is finite.

Combining the fact that the limit exists with sections IV-A, IV-B and IV-C gives the capacity of the FSBC of Theorem 1.